\documentclass[12pt]{article}
\usepackage{graphicx}

\newcommand\pubdate{\today}

\newcommand\pubnumber{NTLP 2011-03}

\def\Title#1{\begin{center} {\Large #1 } \end{center}}
\def\Author#1{\begin{center}{ \sc #1} \end{center}}
\def\Address#1{\begin{center}{ \it #1} \end{center}}

\newcommand\pubblock{\rightline{\begin{tabular}{l} \pubnumber\\
         \pubdate  \end{tabular}}}
\newenvironment{Abstract}{\begin{center}{\bf Abstract}\end{center} \bigskip \begin{quotation}  }{\end{quotation}}
\newenvironment{Presented}{\begin{quotation} \begin{center} 
             PRESENTED AT\end{center}\bigskip 
      \begin{center}\begin{large}}{\end{large}\end{center} \end{quotation}}

\def\beq{\begin{equation}}
\def\eeq#1{\label{#1}\end{equation}}
\def\eeqn{\end{equation}}

\def\beqa{\begin{eqnarray}}
\def\eeqa#1{\label{#1}\end{eqnarray}}
\def\eeqan{\end{eqnarray}}

\let\bar=\overbar

\def\Dslash{\not{\hbox{\kern-4pt $D$}}}
\def\dslash{\not{\hbox{\kern-2pt $\del$}}}

\def\msb{{\bar{\ssstyle M \kern -1pt S}}}

\begin{document}
\begin{titlepage}
\pubblock

\vfill


\Title{Search for Lepton Flavor Violating $\tau$ Decays at $B$-factories}
\vfill
\Author{Yoshiyuki Miyazaki}  
\Address{Nagoya University,
        Furo-cho, Chikusa-ku, Nagoya, Aichi, Japan}
\vfill


\begin{Abstract}
Lepton flavor violations in charged lepton give good signatures 
for the new physics.
We review recent searches for
lepton flavor violation in $\tau^-$ decays
at $B$-factories.
In these searches, optimization for background reduction is
important to obtain high sensitivity.
No evidence for these decays is observed and
90\% confidence level upper limits have been set
on the branching fractions at the $O(10^{-8})$  level.
\end{Abstract}

\vfill

\begin{Presented}
The Ninth International Conference on\\
Flavor Physics and CP Violation\\
(FPCP 2011)\\
Maale Hachamisha, Israel,  May 23--27, 2011
\end{Presented}
\vfill

\end{titlepage}
\def\thefootnote{\fnsymbol{footnote}}
\setcounter{footnote}{0}
%


\section{Introduction}
Lepton-flavor-violating (LFV) decays
of charged leptons are expected to have negligible probability
even including neutrino oscillations in the Standard Model (SM).
The branching fractions of $\tau^-\to\mu^-\gamma$ 
including  SM$+$ neutrino oscillations
are less than $O(10^{-40})$~\cite{model1}.
However, many extensions of SM, such as supersymmetry (SUSY)
and
large extra dimensions,
predict enhanced  LFV decays with branching fractions
close to the current experimental 
sensitivity~\cite{model2,model3,model4,model5,model6}.
With certain combinations of 
new {physics parameters}, 
the branching fractions for 
LFV $\tau$ decays can be as  high as $10^{-7}$,
which is already accessible {in} high-statistics  
{$B$-factory} experiments.
Therefore, an observation of LFV decay will
be a clear signature
for new physics beyond the SM.
$\tau$ leptons are expected to
be coupled strongly with new physics
and have many possible LFV decay modes
due to their large mass.
Therefore, $\tau$ leptons are ideal objects to 
search for the LFV decays.

SUSY, which is the most popular candidate
among New Physics (NP) models, induces naturally LFV at one-loop 
through the scalar lepton mixing.
The $\tau^-\to\ell^-\gamma$ modes,
where $\ell^-$ is either an electron or a muon,
are important  
and have the largest branching fraction 
in the SUSY seesaw model.
The predicted branching fraction of  $\tau^-\to\mu^-\gamma$
is written as 
\begin{equation}
{\cal{B}}(\tau^-\to\mu^-\gamma) = 3.0\times10^{-6}
\times
\left(\frac{\tan\beta}{60}\right)^2
\left(\frac{\rm{1 TeV}}{M_{\rm SUSY}}\right)^{4}
\label{mug_ped}
\end{equation}
where $M_{\rm SUSY}$ is the typical SUSY mass and 
$\tan \beta$ is the ratio of two Higgs vacuum expectation 
values~\cite{mugamma_pred}.
If $M_{\rm SUSY}$ is small and $\tan \beta$ is large,
this decay mode is enhanced up to current experimental sensitivity.

If a typical SUSY mass  is larger than $\sim$ 1 TeV,
processes via one-loop contributions with SUSY particles
are suppressed. 
When scalar leptons  are much heavier than weak scale,
LFV occurs via {a} Higgs-mediated LFV mechanism.
If LFV occurs via {a} Higgs-mediated LFV mechanism, 
{$\tau^-$ leptons can decay
into $\ell^- f_0(980)$, 
through a scalar Higgs 
boson.
{The decays} $\tau^-\to\ell^- \pi^0$, 
$\ell^- \eta$ and $\ell^-\eta'$ 
{are mediated by} a pseudoscalar Higgs boson
{while} $\tau^-\to\ell^-\mu^+\mu^-$ 
{can be mediated through both} 
scalar and pseudoscalar Higgs bosons~\cite{higgsLFV}.

The ratios between theoretically predicted branching fractions of
$\tau^-\rightarrow\mu^-\gamma$,
$\tau^-\rightarrow\mu^-\mu^+\mu^-$,
and
$\tau^-\rightarrow\mu^- e^+e^-$ and 
maximum theoretical branching fraction
of the $\tau^-\rightarrow\mu^-\gamma$  mode
are summarized in Table \ref{ratios}.
Since the ratio of the branching ratios  
allows to discriminate between new physics models,
model-independent searches for various LFV modes 
are very important. 




\begin{table}[h]
\begin{center}
 \begin{tabular}{|c|c|c|c|c|}
\hline
&SUSY+Seesaw& Higgs mediated &Little Higgs & non-universal $Z'$\\
\hline
$\displaystyle \frac{{\cal B}(\tau^-\rightarrow\mu^-\mu^+\mu^-)}{{\cal
  B(\tau^-\rightarrow\mu^-\gamma)}}$ &
  $\sim2\times10^{-3}$&0.06$\sim0.1$&$0.04\sim0.4$&$\sim$20\\
\hline
$\displaystyle \frac{{\cal B}(\tau^-\rightarrow\mu^- e^+e^-)}{{\cal B(\tau^-\rightarrow\mu^+\gamma)}}$&
$\sim1\times10^{-2}$&$\sim1\times10^{-2}$&$0.04\sim0.4$&$\sim20$20\\
\hline
$\displaystyle {\cal
  B}(\tau^-\rightarrow\mu^-\gamma)$
& $<10^{-7}$ & $<10^{-10}$ & $<10^{-10}$ & $<10^{-9}$ \\
\hline
\end{tabular}
\caption{Ratios between the branching fractions
of the $\tau^-\rightarrow\mu^-\gamma$ and
$\tau^-\rightarrow\mu^-\ell^+\ell^-$ modes
and the maximum theoretical branching fraction
of the $\tau^-\rightarrow\mu^-\gamma$  mode
in various new physics models.}
\label{ratios}
\end{center}
\end{table}

\section{KEKB/Belle and PEP-II/BaBar}

The KEKB 
is a $e^+e^-$ asymmetric-energy collider 
operating at the center-of-mass (CM) energy corresponding to the 
$\Upsilon(4S)$ resonance.
KEKB have achieved the world highest peak luminosity of $2.1\times10^{34}$
cm$^{-2}$s$^{-1}$.
Experiments at the energy of $\Upsilon(4S)$ allow searches
for LFV decays with a very high sensitivity since
the cross section of $\tau^+\tau^-$ production
is $\sigma_{\tau\tau}\simeq 0.9$ nb,
close to  that of  $B\bar{B}$ production,
$\sigma_{B\bar{B}}\simeq 1$ nb, and
thus, $B-$factories are also excellent $\tau-$factories.
The Belle detector~\cite{Belle}
operating at the
KEKB $B$-factory~\cite{kekb}
accumulated about $9\times 10^8$ $\tau$ pairs.
Similarly,
the BaBar detector, described in more detail elsewhere~\cite{BaBar},
collected data at the PEP-II asymmetric-energy $e^+e^-$ collider 
that operated at a CM energy of 10.58 GeV. 
Finally, a 557 fb${}^{-1}$ data sample has been accumulated
before the PEP-II collider stopped running.
Both detectors at $B$-factories 
are the multipurpose detectors 
with good track reconstruction 
and particle identification ability.

\section{Analysis Method}

All searches for LFV $\tau$ decays follow  a similar pattern.
We search for $\tau^+\tau^-$ events
in which one $\tau$ (signal side) decays into an LFV mode under study,
while the other $\tau$ (tag side) decays
into one (or three) charged particles
and any number of additional photons and 
neutrinos (for example, see Fig.~\ref{signal_sig}).
To search for exclusive decay modes,
we select low-multiplicity
events with zero net charge,
and separate a signal- and tag-side into two hemispheres using a thrust axis.
The backgrounds in such searches are dominated by
continuum $e^+e^-\to q\overline{q}~(q = u, d, s, c)$, generic $\tau^+\tau^-$, two-photon,
$\mu^+\mu^-$ and Bhabha events.
To obtain good sensitivity,
we optimize the event selection 
using particle identification and
kinematic information for each mode separately. 

\begin{figure}[t]
\begin{center}
\includegraphics[width=19.pc]{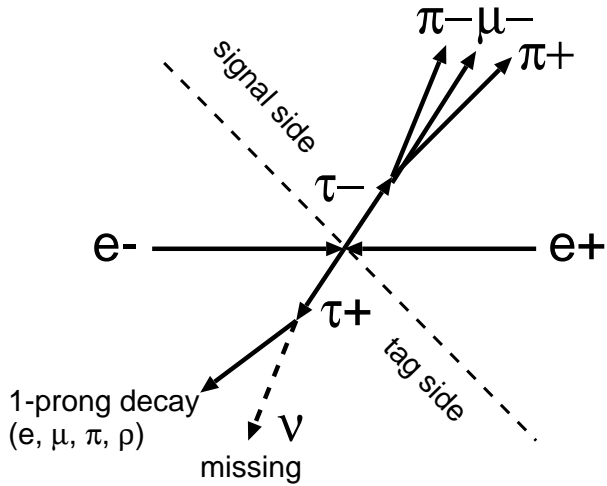}
\caption{
Event signature of LFV $\tau^-$ decay 
in a case of $\tau^-\to\mu^-\pi^+\pi^-$ analysis
}
\label{signal_sig}
\end{center}
\end{figure}

After signal selection criteria are applied, 
signal candidates are examined in the two-dimensional
space of the invariant mass, $m_{\rm {inv}}$, and the difference of
their energy from the beam energy in the center-of-mass (CM) system,
$\Delta E$. A signal event should have $m_{\rm {inv}}$
close to the $\tau$-lepton mass and $\Delta E$ close to 0 GeV.
We blind a region around the signal region
in the $m_{\rm{inv}}-\Delta E$ plane
so as not to bias our choice of selection criteria.
The expected number of background events in the blind region and systematic 
uncertainties are first
evaluated,
and then the blind region is opened and candidate events are counted.
By comparing the expected and observed numbers of events,
we either observe  a LFV $\tau$ decay or set an upper limit
by applying Bayesian, Friedman-Cousins or maximum likelihood approaches.

\section{Results}

\subsection{$\tau^-\to\ell^-\gamma$}

Belle have obtained upper limits 
for the branching fraction
at the 90\% confidence level
${\cal{B}}(\tau^-\rightarrow \mu^-\gamma) < 4.5\times 10^{-8}$
and
${\cal{B}}(\tau^-\rightarrow e^-\gamma) < 1.2\times 10^{-7}$~\cite{cite:lgamma}
using  535 fb$^{-1}$ of data.
The dominant background for these modes
comes from generic $\tau\tau$ events 
where one $\tau$ decays into $\ell\nu\bar{\nu}$
with initial state radiation.
Since  many background events
from $\tau\tau$ with initial state radiation
remain, our
sensitivity is limited.

BaBar updated the search for $\tau^- \to \ell^- \gamma$
using their final data set of 470 fb$^{-1}$ on $\Upsilon(4S)$,
31 fb$^{-1}$ on $\Upsilon(3S)$ and 15 fb$^{-1}$ on $\Upsilon(2S)$,
which corresponds to $(963 \pm 7) \times 10^6$ $\tau$ decays.
In this analysis, new kinematic cuts and a neural-net discriminator
were applied. 
The $m_{\rm inv}$-$\mit\Delta E$ distributions are shown in 
Fig.~\ref{fig:babarlg}.
The efficiency was 6.1 and 3.9\% 
for the $\tau^- \to \mu^- \gamma$ and 
$e^- \gamma$ modes, respectively. 
The number of expected background events was $3.6\pm0.7$ and 
$1.6\pm0.4$ while observed number of data in the signal region
are 2 and 0 events for the $\tau^- \to \mu^- \gamma$ and 
$e^- \gamma$ modes, respectively. 
BaBar set the upper limits of
branching fraction to be
$<4.4\times10^{-8}$ and
$<3.3\times10^{-8}$ for 
the $\tau^- \to \mu^- \gamma$  and  $\tau^- \to e^- \gamma$ modes 
at the 90\% CL, repsectively~\cite{babarlg}.
\begin{figure}[h]
\centerline{
\includegraphics[width=13pc]{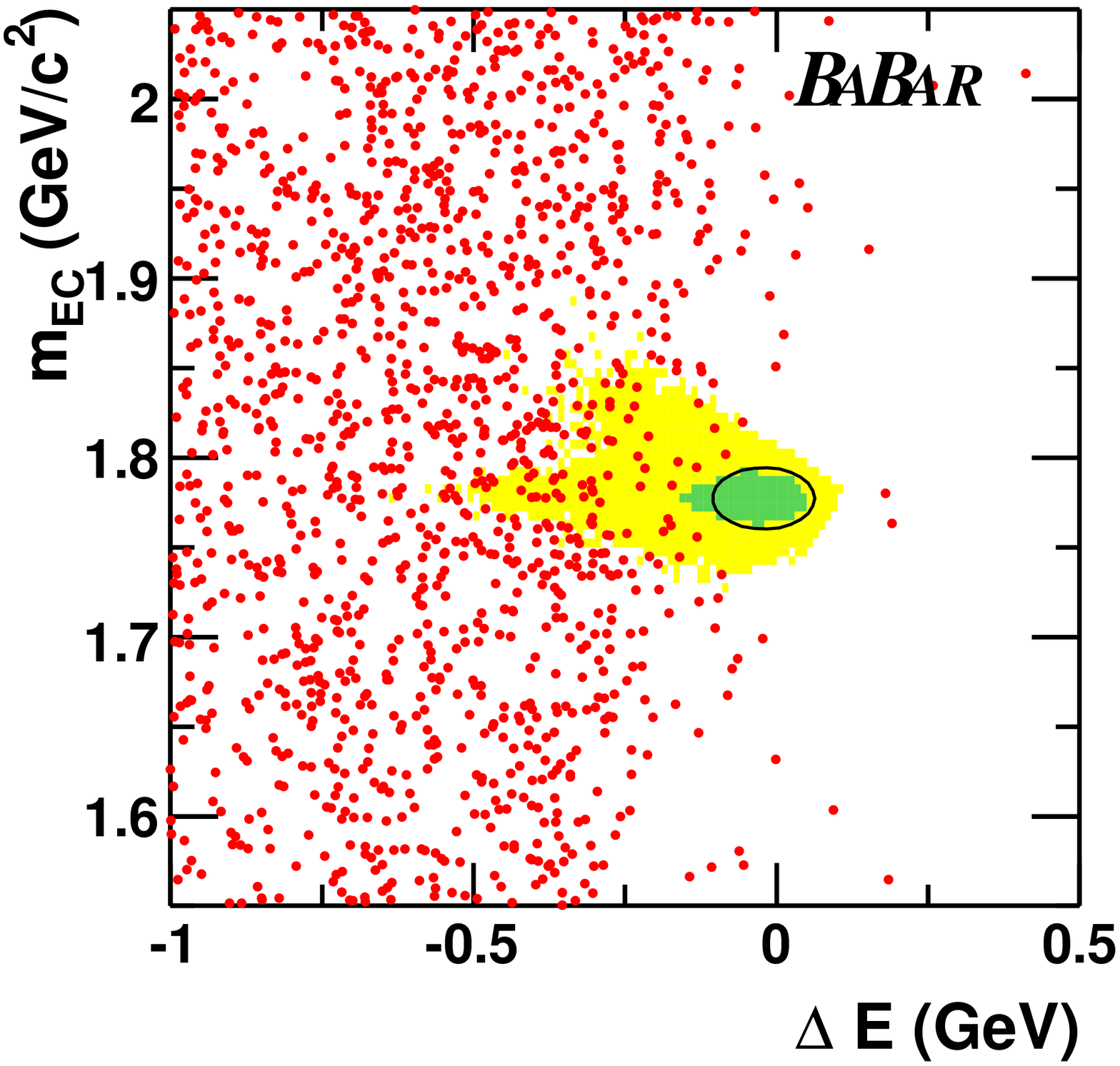}
\includegraphics[width=13pc]{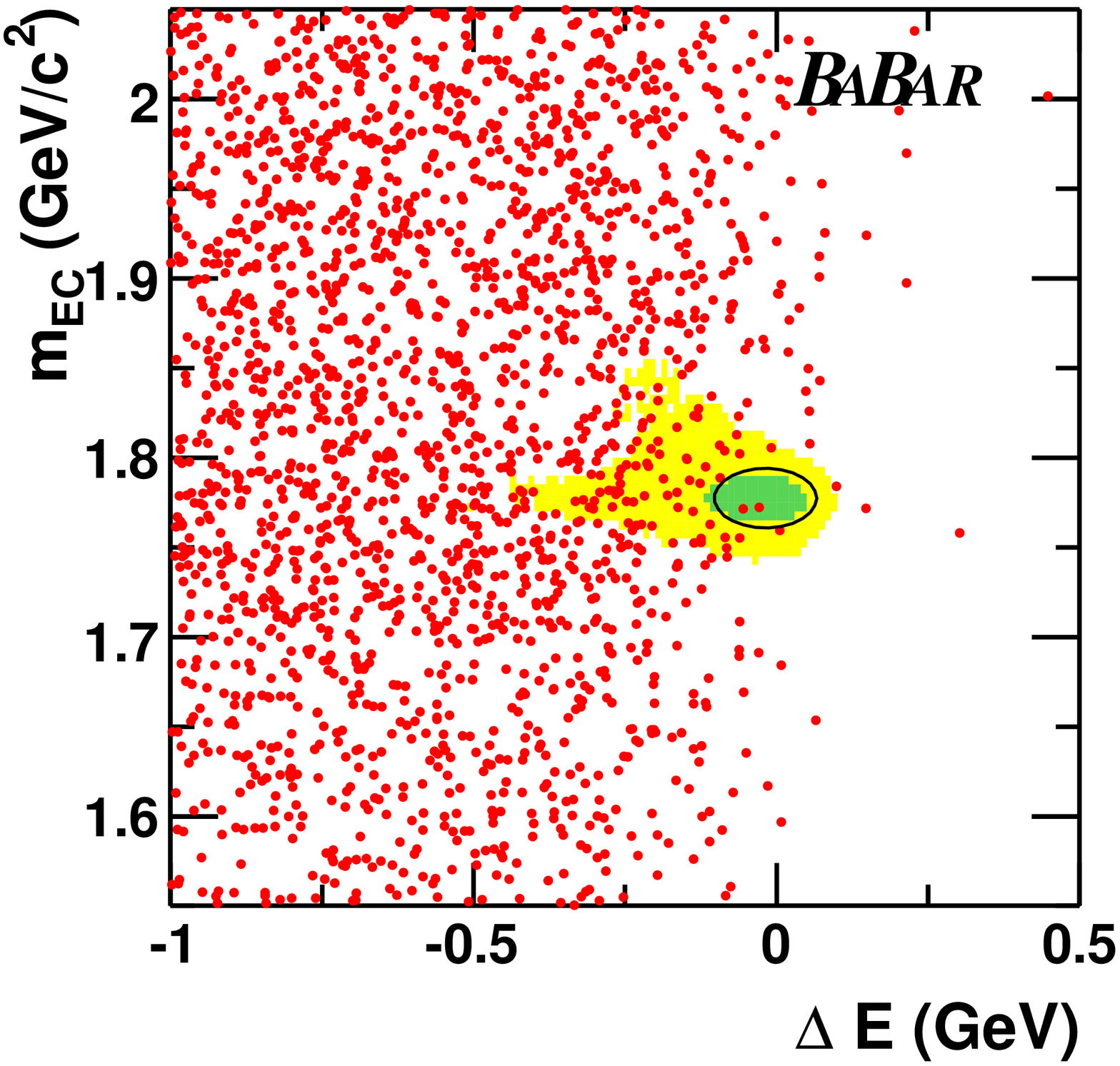}
}
\caption{$m_{\rm inv}$-$\mit\Delta E$ distributions for $\tau^- \to e^- \gamma$ 
and $\tau^- \to \mu^- \gamma$ from the BaBar analysis.
Data are
shown as dots and contours containing 90\% (50\%) of signal MC events 
are shown as yellow- 
(green-) shaded regions.
The elliptical signal
{regions}
shown by a solid curve
are used for evaluating the signal yield.}
\label{fig:babarlg}
\end{figure}

\subsection{$\tau^-\to\ell^-\eta, \ell^-\eta', \ell^-\pi^0$}

Belle and BaBar have published
the results of the search for the $\tau^-$ decays into a lepton
and a neutral pseudoscalar $(\pi^0, \eta, \eta')$ 
using  around 400fb${}^{-1}$ of data, 
and have set the range of the upper limits of $(0.8-2.4)\times10^{-7}$
at 90\% CL.~\cite{Belle:leta,BaBar:leta}

Belle updated a search
for these modes using 901 fb$^{-1}$ of data.
By introducing new event selections with a neural-net discriminator
and studying the background components in detail, 
they obtain larger efficiencies than previous analysis 
in a factor of around 1.5 in the averages, and
the expected numbers of the background events in the signal
region are achieved to suppress less than one events for each mode. 
One event is found in the signal region
for the $\tau^-\rightarrow e^-\eta(\rightarrow\gamma\gamma)$ mode
while no event is observed in other modes (see Fig.~\ref{fig:lp0}).
Therefore, no evidence for these decays is observed and
Belle sets preliminary 90\% confidence level upper limits
on the branching fractions between
$(2.2-4.4)\times10^{-8}$.

\begin{figure}[h]
\begin{center}
\includegraphics[width=25pc]{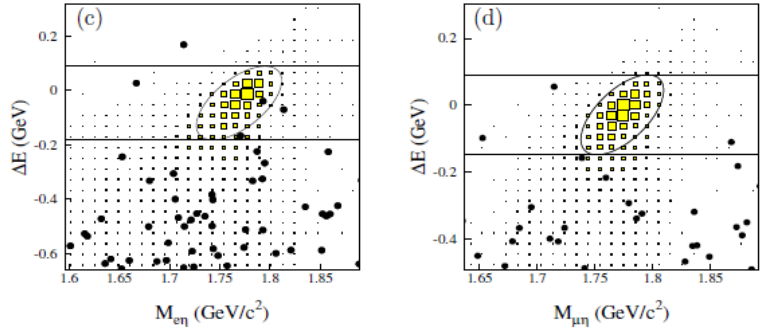}
\caption{$m_{\rm inv}$-$\mit\Delta E$ distributions for 
the $\tau^- \to e^- \eta(\to\gamma\gamma)$ (left)
and $\tau^- \to \mu^-\eta(\to\gamma\gamma)$ (right) from the Belle analysis.
Data and signal MC events are
shown as dots and histogram.
The elliptical signal
{regions}
shown by a solid curve
are used for evaluating the signal yield.
}
\label{fig:lp0}
\end{center}
\end{figure}

\subsection{$\tau^-\to\ell V^0$}

Previously, Belle obtained
90\% confidence level (C.L.) upper limits
{on} branching 
{fractions of these decays using}
543 fb${}^{-1}$ of data,
and these results were
in the range (5.8$-$18)~$\times~10^{-8}$~\cite{lv0_belle}.
The BaBar collaboration
has
also
published {90\% C.L.
upper limits}
in the range (2.6$-$19)~$\times~10^{-8}$
using 451 fb${}^{-1}$ of data~\cite{lv0_babar}
{for all $\tau^-\to\ell^- V^0$ decays 
{except for} 
{$\tau^-\to\ell^-\omega$} 
for which 384 fb${}^{-1}$ of data were used~\cite{lomega_babar}.}

Belle updates an search
for these modes based on a data sample of
854~fb$^{-1}$ of data.
To improvement better results than previous analysis,
we use {a} larger data sample and apply improved rejections
of specific backgrounds, 
such as di-baryon production in the continuum 
for the $\tau^-\to\mu^- V^0$ modes, 
and $\tau^-\to h^-\pi^0\nu_{\tau}$ decays with a photon conversion
for the $\tau^-\to e^-V^0$ modes.} 
Since no evidence for a signal after event selections
is {found, we set the} following 90\% C.L. upper limits 
{on the branching fractions:} 
${\cal{B}}(\tau^-\rightarrow e^-V^0)
 < (1.8-4.8)\times 10^{-8}$
and 
${\cal{B}}(\tau^-\rightarrow \mu^- V^0)< (1.2-8.4)\times 10^{-8}$~\cite{lv0_belle854} (for example, see Fig.~\ref{fig:lv0}).
{These results improve {upon} 
our previously published upper limits
by factors {of} up to 5.7.

\begin{figure}[h]
\begin{center}
\includegraphics[width=14pc]{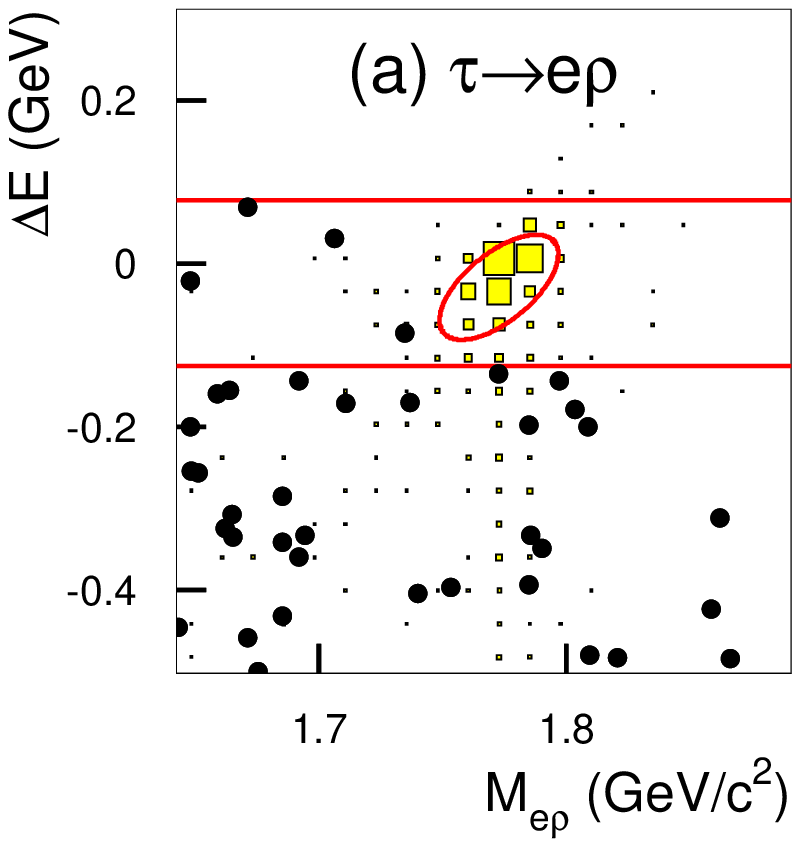}
\includegraphics[width=14pc]{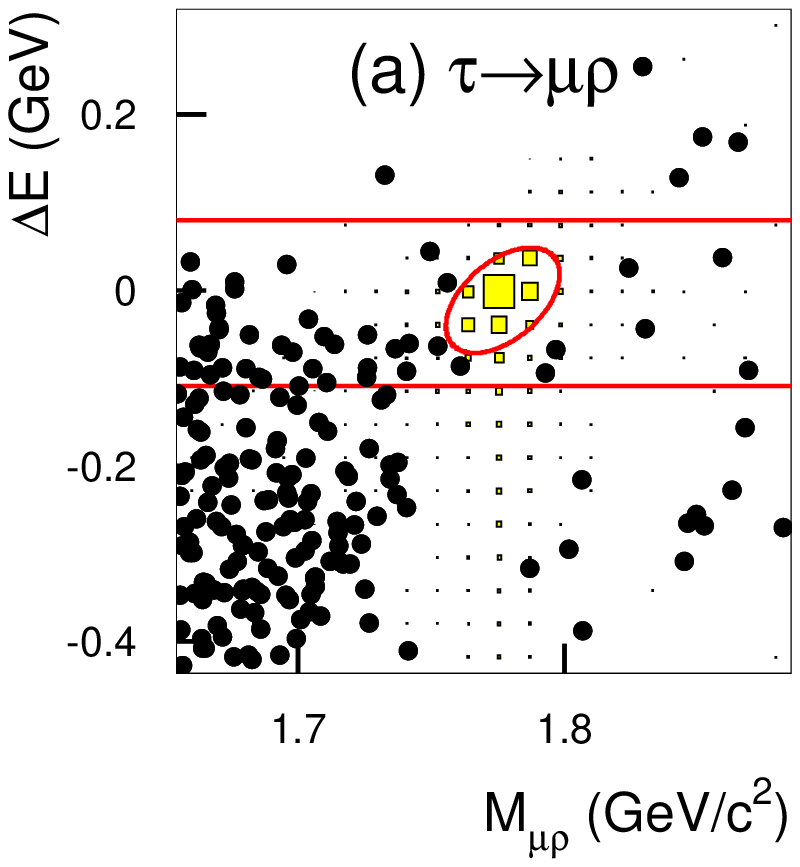}
\caption{$m_{\rm inv}$-$\mit\Delta E$ distributions for 
the $\tau^- \to e^- \rho$ (left)
and $\tau^- \to \mu^-\rho$ (right) modes from the Belle analysis.
Data and signal MC events are
shown as dots and histogram.The elliptical signal
{regions}
shown by a solid curve
are used for evaluating the signal yield.}
\label{fig:lv0}
\end{center}
\end{figure}

\subsection{$\tau\to\ell hh'$}

Belle and BaBar 
have 
also searched for various $\ell hh'$
(where $h,~h'= \pi^{\pm}$ or $K^{\pm}$) modes 
including lepton flavor 
and 
lepton number violation 
($\tau^-\to\ell^-h^-h'^+$  and $\tau^-\to\ell^+h^-h'^-$)
with the range of upper limits:
 (7$-$48)~$\times~10^{-8}$~\cite{lhh_babar} 
and (5.8$-$18)~$\times~10^{-8}$~\cite{lhh_belle}
using 221 fb${}^{-1}$ and 543 fb${}^{-1}$  
of data, respectively.

Belle recently updated a search
for 
these modes
using 854 fb$^{-1}$ of data.
Basically, Belle applies similar event selections 
to $\tau^-\to\ell^- V^0$ analysis due to same signature as $\tau\to\ell V^0(\to hh')$ in the final state.
However, 
the remaining background events for these modes are larger than
 $\tau^-\to\ell^-  V^0$ modes 
since there is no requirements to reconstruct $V^0$ mesons.
Therefore, 
Belle applies additional tighter selection as missing informations, for example.
After event selection, Belle observes one events 
in the signal region 
for the $\tau^-\to\mu^+\pi^- \pi^-$ and $\tau^-\to e^+\pi^- K^-$ modes
while no events are found for the other modes.
In each case, the number 
of events observed in the signal {region}
{is} 
consistent with
{the} expected number of background events.
Therefore, no evidence for these decays is observed,
and we set 
preliminary
upper limits
on the branching fractions at 90\% C.L.:
${\cal{B}}(\tau\rightarrow ehh')
< (2.0-3.7)\times 10^{-8}$
and
${\cal{B}}(\tau\rightarrow \mu hh')
< (2.1-8.6)\times 10^{-8}$.
These results improve {upon} 
previously Belle published upper limits
by factors {of} around 1.8 on the average.

\section{Future Prospect}

LFV sensitivity depends on the
remaining background level.
For the $\tau^-\to\mu^-\gamma$ mode,
there is large remaining background 
from generic $\tau^+\tau^-$ events with initial state radiation.
In this case,
the expected branching fraction of 
$\tau^-\to\mu^-\gamma$ is scaled as $1/\sqrt{\cal{L}}$.
On the other hand,
the remaining background events for
the $\tau^-\to\ell^-\ell'^+\ell''^-$ and $\ell^-$+meson modes
are expected to be negligible at 10 ab${}^{-1}$.
Therefore,
the expected branching fractions of 
these modes  are  linearly proportional to luminosity
from current upper limits.
Figure~\ref{fig} shows the history of the obtained UL of the 
branching fractions
as a function of the integrated luminosity, 
as well as the expected
sensitivity extrapolating from the results.
The upgraded $B$-factories, Belle-II and Super$B$ experiments,
are planned to 
collect more than 10-times larger
luminosity than the  current one.
Therefore, 
the expected branching fraction of 
$\tau^-\to\mu^-\gamma$
at the Super $B-$factory is $O(10^{-(8\sim9)})$
while the
expected branching fractions of 
$\tau^-\to\ell^-\ell'^+\ell''^-$ and $\ell^-$+meson 
are $O(10^{-(9\sim10})$.

\begin{figure}[h]
\begin{center} 
\includegraphics[width=18pc]{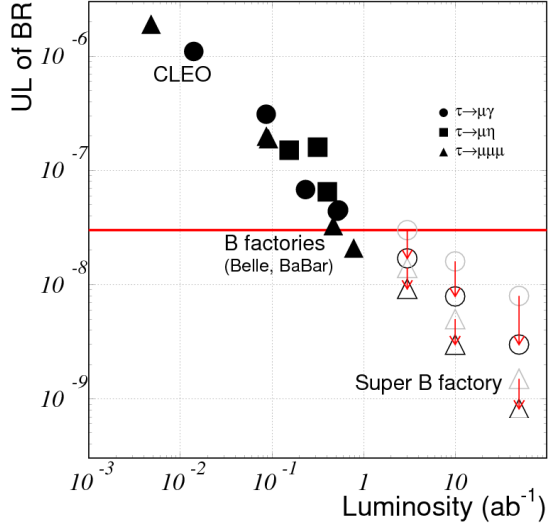}
\caption{
Branching fraction of LFV decay 
as a function of the integrated luminosity
as well as the expected
sensitivity extrapolating from the current results.}
\label{fig}
\end{center}
\end{figure}

\section{Summary}

We have searched for
all major modes of lepton-flavor-violating
$\tau$ decays using
$>10^9$ $\tau$ pairs of data collected
at the $B$-factories as 
the Belle detector at the
KEKB 
collider
and the BaBar detector at the
PEP-II 
collider.
No evidence for these decays is observed and
we set 90\% confidence level upper limits
on the branching fractions at the {$O(10^{-8})$ 
level, shown in Fig.~\ref{fig:LFV_summary} and Table~\ref{LFV_summary}.
These more stringent upper limits can be used
to constrain the space of parameters in various models beyond
the SM.

\begin{figure}[h]
\begin{center}
\includegraphics[width=30pc]{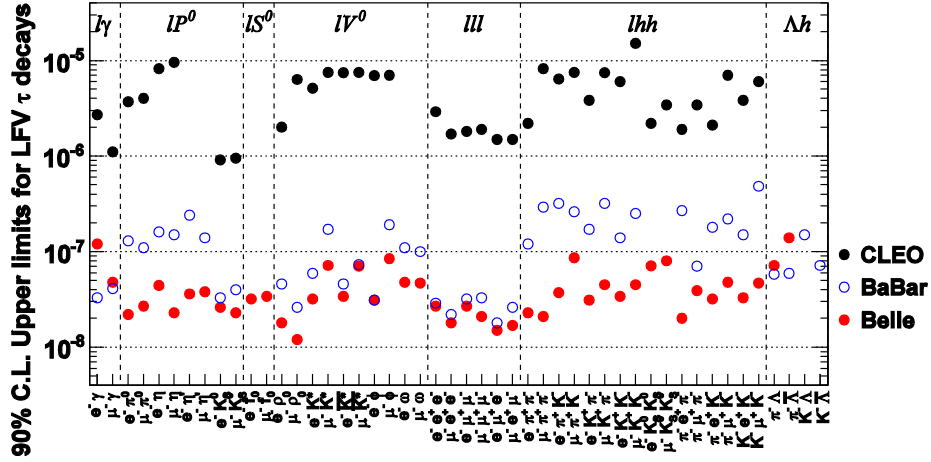}
\caption{
Upper limits of branching fraction of lepton-flavor-violating  $\tau$ decay
from Belle, BaBar and CLEO.}
\label{fig:LFV_summary}
\end{center}
\end{figure}

\begin{table*}[t]
\begin{center}
\begin{tabular}{|c|c|c|c|c|}\hline
& \multicolumn{2}{c|}{Belle} &  \multicolumn{2}{c|}{BaBar} \\ \cline{2-5}
$\tau^-$ decay mode &  $\mathcal{B}, 10^{-8}$ & \# of $\tau^+\tau^-$
& $\mathcal{B}, 10^{-8}$  & \# of $\tau^+\tau^-$\\ \hline
$\mu^- \gamma$ & 4.5~\cite{cite:lgamma}  & 492M  &
 4.4~\cite{babarlg} & 482M  \\
$e^- \gamma$ & 12~\cite{cite:lgamma} & 492M &
 3.3\cite{babarlg} & 482M   \\
$\mu^- \eta, \mu^- \eta', \mu^- \pi^0$  & 6.5-13
&
 369M
& 11-20~\cite{BaBar:leta} & 312M \\
$e^- \eta, e^- \eta', e^- \pi^0$  & 8.0-16 & 369M
& 14-26~\cite{BaBar:leta} & 312M \\
$\ell^-\ell'^-\ell''^+$ & 1.5-2.7~\cite{new_lll} & 719M  
& 1.8-3.3~\cite{3l_babar}  & 430M  \\
$\mu^-hh'$ & 2.1-8.6 & 782M &
 7-44~\cite{lhh_babar}  & 203M  \\
$e^-hh'$ & 2.0-3.7 & 782M &
12-32~\cite{lhh_babar}  & 203M  \\
$\ell^- f_0(980)(\to\pi^+\pi^-)$ & 3.2-3.4~\cite{cite:lf0} & 617M & -- & -- \\
$\mu^- V^0$ & 1.2-8.4~\cite{lv0_belle854} & 782M & 8-18~\cite{lv0_babar,lomega_babar} & 414M \\
$e^- V^0$ & 1.8-4.8~\cite{lv0_belle854} & 782M & 3.1-5.6~\cite{lv0_babar,lomega_babar}
& 414M \\
$\mu^- K^0_S$ & 2.3~\cite{cite:lks_new} &617M  & 4.0~\cite{cite:babar_lks} & 431M \\
$e^- K^0_S$ & 2.6~\cite{cite:lks_new} & 617M & 3.3~\cite{cite:babar_lks} & 431M \\
$\mu^- K^0_SK^0_S$ & 8.0~\cite{cite:lks_new} & 617M & -- & -- \\
$e^- K^0_SK^0_S$ & 7.1~\cite{cite:lks_new} & 617M & -- & -- \\
\hline
\end{tabular}\\[2pt]
\caption{Summary for upper limits of branching fraction of lepton-flavor-violating  $\tau$ decay}
\label{LFV_summary}
\end{center}
\end{table*}

\end{document}